%% file: paper_v2.tex
\documentclass[3p,times]{elsarticle}
 
\usepackage{ecrc}
 
\volume{00}
 
\firstpage{1}
 
\journalname{Nuclear Instruments and Methods in Physics Research Section B}
 
\runauth{Chang et al.}
 
\jid{nimprb}
 
\jnltitlelogo{NIMPR-B}

\usepackage{lineno,hyperref} 
\modulolinenumbers[5] 
 
\usepackage{mathtools}
\usepackage[usenames, dvipsnames]{xcolor}
\usepackage{subcaption}

\usepackage{tikz}
\usepackage{tikz-qtree}
\usepackage{tkz-berge}
\usepackage{sansmath}
\usetikzlibrary{shadings,intersections}
\usetikzlibrary{shapes.geometric, arrows.meta}
\usepackage{tikz-3dplot}
\usetikzlibrary{calc,3d,decorations.markings,backgrounds,positioning,intersections,shapes,quotes,angles}
\usepackage{nicefrac}
\usepackage{caption}
\usepackage{subcaption}
\usepackage{graphicx}
\usepackage{caption,booktabs}
\usepackage{amsmath}
\usepackage{amstext}

\usepackage[]{hyperref}
\hypersetup{
    colorlinks,%
    citecolor=blue,%
    filecolor=black,%
    linkcolor=blue,%
    urlcolor=black
}
 
\usepackage{cleveref}
\usepackage{tabularx}
\usepackage{multicol}
 
\biboptions{sort&compress}
 
\usepackage{scrextend}
\begin{document}
 
\begin{frontmatter} 
\title{Monte Carlo modeling of low-energy electron-induced secondary electron emission yields in micro-architected boron nitride surfaces} 

\author[UCLA_MSE]{Hsing-Yin Chang}
\ead{irischang@ucla.edu}
\author[UCLA_MSE]{Andrew Alvarado}
\author[UCLA_MSE]{Trey Weber}
\author[UCLA_MSE,UCLA_MAE]{and Jaime Marian}
 
\address[UCLA_MSE]{Department of Materials Science and Engineering, University of California, Los Angeles, CA 90095, USA}
\address[UCLA_MAE]{Department of Mechanical and Aerospace Engineering, University of California, Los Angeles, CA 90095, USA}
 
\begin{abstract}
Surface erosion and secondary electron emission (SEE) have been identified as the most critical life-limiting factors in channel walls of Hall-effect thrusters for space propulsion. Recent wall concepts based on micro-architected surfaces have been proposed to mitigate surface erosion and SEE. The idea behind these designs is to take advantage of very-high surface-to-volume ratios to reduce SEE and ion erosion by internal trapping and redeposition. This has resulted in renewed interest to study electron-electron processes in relevant thruster wall materials.
In this work, we present calculations of SEE yields in micro-porous hexagonal BN surfaces using stochastic simulations of electron-material interactions in discretized surface geometries. Our model consists of two complementary parts. First we study SEE as a function of primary electron energy and incidence angle in flat surfaces using Monte Carlo simulations of electron multi-scattering processes. The results are then used to represent the response function of discrete surface elements to individual electron rays generated using a ray-tracing Monte Carlo model. We find that micro-porous surfaces result in SEE yield reductions of over 50\%  in the energy range experienced in Hall thrusters. This points to the suitability of these micro-architected surface concepts to mitigate SEE-related issues in compact electric propulsion devices. \end{abstract}
 

\end{frontmatter}
 
\section{Introduction}
Advances in electrode, chamber, and structural material technology will enable breakthroughs in future generations of electric propulsion and pulsed power (EP $\&$ PP) concepts \cite{martinez1998spacecraft,goebel2008fundamentals}. Although significant advances have been achieved during the past few decades, much of the progress has relied on empirical development of materials through experimentation and trial-and-error approaches \cite{raitses2011effect,patino2015analysis}. 
Materials and channel wall designs are sought by optimizing performance against  weight, power density, and cost. Under extreme operating environments, the discharge channels and cathode insulators used in Hall thrusters require a very demanding set of properties. Typically, the range of materials of interest in EP $\&$ PP include refractory metals, such as W, Mo, and their alloys, ceramic composites, such as BN and $\rm Al_{2}O_{3}$, high-strength copper alloys, and carbon-carbon composites \cite{levchenko2018recent}. These classes of materials possess great mechanical strength, thermal shock resistance, refractoriness and machinability. 
Surface erosion and secondary electron emission (SEE) from the channel walls have been identified as among the most critical life-limiting factors for Hall thrusters \cite{clauss1997preliminary}. To mitigate surface erosion and SEE, a promising route is to modify the surface roughness. Recently, demonstration designs have been developed, including surface architectures based on metal micro-spears, micro-nodules and micro-velvets \cite{raitses2006operation,ye2013suppression,yang2015nanostructured,jin2017secondary,huerta2018secondary}. See reviews on the topic for further information \cite{raitses2006measurements,jolivet2000effects,campanell2012general,campanell2012absence,mazouffre2016electric,levchenko2018recent}. 

The main objective of our work is to develop reliable physical models of SEE, to be applied to the calculation of effective SEE yields in micro-architected surfaces for space propulsion thrusters. The methodology relies on two distinct but complementary elements. The first is an experimentally validated theoretical model of electron scattering in solids. This model is built as a transport Monte Carlo simulator of individual electron trajectories in a solid, capturing the pertinent scattering mechanisms in terms of an interaction differential cross section that is integrated across the relevant energy and angular ranges. These cross sections reflect different elastic and inelastic (e.g. core electrons, valence electrons, polarons, plasmons, etc) scattering mechanisms in each material, and is formulated according to the best available physics. The second element of the methodology is a discretization procedure to represent arbitrary surface geometries in terms of discrete boundary elements. Both modules (physics and geometry) are coupled by way of a raytracing algorithm that captures the intersection of primary electron rays with different boundary elements. The material considered here is BN (used in current thrusters thanks to its low mass density, low thermal expansion, and highly dielectric properties). The procedure is described in two papers from our group focused on W \cite{chang2018calculation,alvarado2018monte}.
 
The structure of the paper is as follows. First, the theoretical models employed to study electron interaction with insulators is described. Subsequently, these models are validated experimentally for SiO$_2$ and hexagonal BN. The resulting SEE yield and emitted energy distributions for BN are then used as response functions in ray-tracing Monte Carlo simulations of SEE in discretized complex foams.  We finalize with a summary of the main findings and the conclusions and acknowledgments.

\section{Theory and Methods}

\subsection{Electron-Insulator Interaction Model}
The energy loss mechanisms for internal secondary electrons differ between metals and insulators. In metals, the internal secondary electrons lose energy through interactions with conduction electrons, lattice vibrations --known as \emph{polarons}--, and defects. In order to escape, the kinetic energy of an internal secondary electron must be above the energy barrier --taken to be equal to the Fermi level $E_{F}$ plus work function $\Phi$ (typically $>$10 eV)--when it reaches the surface. This large threshold escape energy, as well as the high collision probability due to the large number of conduction electrons, result in the low SEE yields (usually $<$1) found in metals. 
Conversely, due to the low number of conduction electrons in insulators, internal secondary electrons lose energy primarily through the excitation of valence electrons into the conduction band. This prevents secondary electrons with kinetic energies below the bandgap energy from participating in such electron-electron collisions, significantly increasing their mean escape depths compared to that of metals \cite{grais1982study}. The mean escape depth for insulators ranges from 10-50 nm, compared to 0.5-1.5 nm for conductors \cite{seiler1983secondary}. Consequently,  SEE yields in insulators are typically substantially higher than in metals. In some reported cases, the SEE yields of certain insulators can exceed those of standard conductors by as much as a factor of 20 \cite{seiler1983secondary}. 

When the energy of the primary electron beam is considerably higher than the bandgap energy $E_{g}$ of an insulator material, the elementary scattering processes are essentially those encountered for metals. However, at  low energies, several new aspects of the electron-material interaction processes become important. For example, at a low-energy electron within an insulator can locally distort the lattice and greatly reduce its mobility by forming a polaron. As well, defects and impurities can also act as traps.  In any case, a distinctive feature of secondary electron emission in insulators is the buildup of charge, such that subsequent SEE must be considered in the context of the existence of an internal electric field.
In the present work, our calculations consider the following electron-material interaction processes: (i) Mott's theory for electron-atom interactions, (ii) Ritchie's theory for electron-electron interactions, (iii) Fr{\"o}hlich's perturbation theory for electron-phonon interactions, and (iv) Ganachaud and Mokrani's semi-empirical model for electron-polaron interaction. 

\subsection{Elastic Scattering}
Due to the large mass difference of electrons and atomic nuclei, electron-atom collisions can be approximated as being perfectly elastic. 
A commonly used elastic scattering cross-section is the screened Rutherford cross-section, which has a convenient analytical form and is straightforward to implement in a Monte Carlo calculation. However, the screened Rutherford cross-section can be applied only to high-energy electrons and solids with a low atomic number. An alternative to the screened Rutherford cross-section is the relativistic partial wave expansion method of the Mott scattering cross-section.
In Mott's theory \cite{mott1995scattering}, the differential elastic scattering cross section (DESCS) with respect to solid angle $\Omega$ can be calculated as
\begin{equation}
\frac{d\sigma_{el}}{d\Omega} = |f(\theta)|^{2}+|g(\theta)|^{2}
\end{equation}
where $|f(\theta)|$ and $|g(\theta)|$ are the direct and indirect scattering amplitudes, respectively, given by
\begin{equation}
\begin{aligned}
f(\theta) & = \frac{1}{2iK}\sum_{l=0}^{\infty}\{(l+1)[\exp(2i\delta_{l}^{+})-1] \\
& +l[\exp(2i\delta_{l}^{-})-1]\}P_{l}(\cos\theta)
\end{aligned}
\end{equation}
\begin{equation}
g(\theta) =  \frac{1}{2iK}\sum_{l=1}^{\infty}[\exp(2i\delta_{l}^{-})-\exp(2i\delta_{l}^{+})]P_{l}^{1}(\cos\theta).
\end{equation}
In these equations, $K$ is the momentum of the electron, $E$ the total energy, $m$ the electron mass, $c$ the speed of light, $P_{l}(\cos \theta)$ the Legendre polynomials, and $P_{l}^{1}(\cos \theta)$ the first-order associated Legendre polynomials:
\begin{equation}
P_{l}^{1}(x) = \left(1-x^{2}\right)^{\frac{1}{2}}\frac{dP_{l}(x)}{dx}.
\end{equation}
The phase shifts $\delta_{l}^{\pm}$ can be computed by using the equation
\begin{equation}
\tan \delta_{l}^{\pm} = \frac{Kj_{l+1}(Kr)-j_{l}(Kr)[\xi \tan \phi_{l}^{\pm}+(1+l+k^{\pm})/r]}{Kn_{l+1}(Kr)-n_{l}(Kr)[\xi \tan \phi_{l}^{\pm}+(1+l+k^{\pm})/r]}
\end{equation}
where
\begin{equation}
\xi = \frac{E+mc^{2}}{\hslash c}.
\end{equation}
$k^{+} = -l-1$, $k^{-} = l$, $j_{l}$ are the regular spherical Bessel functions, $n_{l}$ the irregular spherical Bessel functions and
\begin{equation}
\phi_{l}^{\pm} = \lim_{r\to\infty}\phi_{l}^{\pm}(r)
\end{equation}
where $\phi_{l}^{\pm}$ is the solution of the Dirac's equation which can be reduced, as shown by Lin, Sherman, and Percus \cite{lin1963elastic} and by Bunyan and Schonfelder \cite{bunyan1965polarization}, to the first-order differential equation
\begin{equation}
\frac{d\phi_{l}^{\pm}(r)}{dr} = \frac{k^{\pm}}{r}\sin[2\phi_{l}^{\pm}(r)]-\frac{mc^{2}}{\hslash c}\cos[2\phi_{l}^{\pm}(r)]+\frac{E-V(r)}{\hslash c}
\end{equation}
with $V(r)$ being the electron-atom potential.

In the present calculation, the atomic differential cross-sections for elastic scattering are extracted from the NIST \emph{Electron Elastic-Scattering Cross-Section Database} \cite{jablonski2010nist} ranging from 0$^{\circ}$ to 180$^{\circ}$ for 24 incident energies between 50 eV and 1 keV (in increments of 10 eV from 50 to 100 eV, and in increments of 50 eV from 100 eV to 1 keV) and put into a data file in tabulated form. The elastic differential cross-sections for energies and angles other than those in the table can be calculated by linear interpolation accurate to two decimal places.

\begin{figure}[htbp] 
\begin{center}
\includegraphics{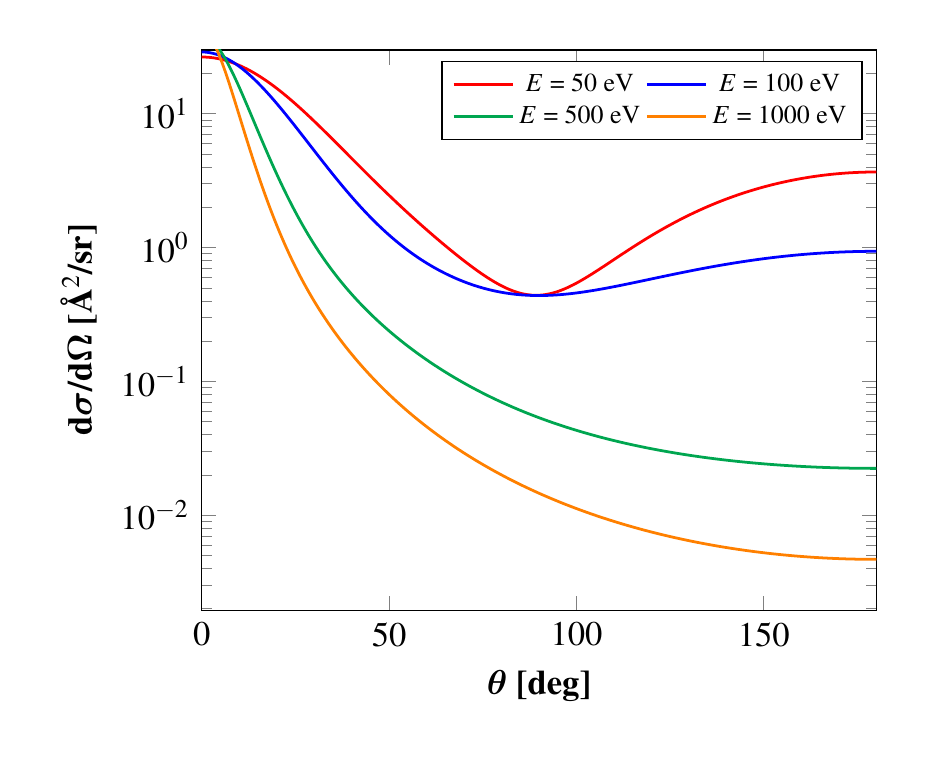}
\caption{Values of the DESCS of 50, 100, and 500 1000 eV electrons scattered by $h$-BN as a function of the scattering angle.}
\label{fig:figure1}
\end{center}
\end{figure}

For compounds, the DESCS can be approximated by using the additivity rule, as the sum of the atomic DESCS of all atoms in the molecule. With the DESCSs of the single elements B and N, the electron-molecule DESCS is obtained for hexagonal boron nitride ($h$-BN) through
\begin{equation}
\bigg(\frac{d\sigma_{el}(E,\vartheta)}{d\Omega}\bigg)_{\rm BN} =\bigg(\frac{d\sigma_{el}(E,\vartheta)}{d\Omega}\bigg)_{\rm B}+\bigg(\frac{d\sigma_{el}(E,\vartheta)}{d\Omega}\bigg)_{\rm N}.
\end{equation}
Figure~\ref{fig:figure1} shows the DESCS of 50, 100, 500 and 1000-eV electrons scattered by $h$-BN as a function of the scattering angle.

The total elastic scattering cross section for electron-molecule interaction can then be calculated as:
\begin{equation}
\sigma_{el}(E) = \frac{\lambda_{el}^{-1}}{N} = 2\pi \int_{0}^{\pi}\frac{d\sigma_{el}(E,\vartheta)}{d\Omega}\sin\vartheta d\vartheta \\
\end{equation}
where $N$ is the number of molecules per unit volume in the target, $\lambda_{el}$ is the elastic mean free path and $\vartheta$ is the polar scattering angle.

At low energies, the elastic mean free path calculated using the phase-shift method becomes very small --on the order of 0.1 nm--, which is much smaller than the interatomic distances and therefore unphysical. This is probably a consequence of using a rigid static potential, as well as neglecting dynamic effects such as that associated with the polarization of the electron cloud. In any case, electrons undergoing elastic collisions become confined in a very small region of space, effectively becoming trapped. Consequently, these processes do not contribute appreciably to secondary electron emission, but may slow down calculations significantly. To address this issue, in this work the elastic cross-sections calculated by the phase-shift method are multiplied by a cut-off function whose role is to gradually decrease the importance of the elastic effect at very low energies. This function must tend towards unity as the energy increases so that the behavior of the static potential used in the phase-shift method is recovered. The choice of function is of course not unique but an expression of the form
\begin{equation}
R_{c}(E) = \tan[\alpha_{c}(E/E_{g})^{2}]
\end{equation}
is seen to behave well for this purpose ($\alpha_{c}$ is a dimensionless parameter and $E_g$ is bandgap). 

The angle of electron scattering at a certain step can be expressed by a random number
\begin{equation}
R = P_{el}(E,\theta) = \frac{1}{\sigma_{el}}\int_{0}^{\theta}\frac{d\sigma_{el}(E,\vartheta)}{d\Omega} \sin\vartheta d\vartheta.
\end{equation}

\subsection{Inelastic Scattering} 
Inelastic scattering is characterized by the energy loss function (ELF) $\operatorname{Im}\big[-\frac{1}{\varepsilon(\textbf{q},\Delta E)}\big]$, where $\textbf{q}$ is the momentum transfer and $\Delta E$ is the energy loss. The dielectric function $\varepsilon(\textbf{q},\Delta E)$ in the ELF reflects the response of a solid to an external electromagnetic perturbation. Due to the difficulties of determining the energy loss function experimentally, Ritchie and Howie suggested an approximate function from the optical dielectric constants by fitting the measured optical data into a finite sum of Drude-Lindhard model functions in the optical limit ($\textbf{q} = 0$) \cite{ritchie1977electron}. In this fashion one can extend the explicit formula to the required $\operatorname{Im}\big[-\frac{1}{\varepsilon(0,\Delta E)}\big]$ for finite $\textbf{q}$-values.

The ELF of the material is parametrized in terms of an expansion of Drude-Lindhard-type oscillators at the optical limit with $N$-term analytic form, which are directly obtained from the features observed in the reflection electron energy loss spectroscopy (REELS) spectrum:
\begin{equation}
\operatorname{Im}\bigg[-\frac{1}{\varepsilon(\textbf{q},\Delta E)}\bigg] = \sum_{i=1}^{n}\frac{A_{i}\gamma_{i}\Delta E}{((\hslash \omega_{0i\textbf{q}})^{2}-\Delta E^{2})^{2}+\gamma_{i}^{2}\Delta E^{2}} \times \theta(\Delta E-E_{g}),
\end{equation}
where
\begin{equation}
\hslash \omega_{0i\textbf{q}} = \hslash \omega_{0i} + \zeta \frac{\hslash^{2}\textbf{q}^{2}}{2m}
\label{hcortada}
\end{equation}
and $A_{i}$, $\gamma_{i}$, and $\hslash \omega_{0i\textbf{q}}$ are the oscillator strength, the damping coefficient, and the excitation energy of the $i$th oscillator, respectively. The step function $\theta(\Delta E-E_{g})$ simulates a bandgap for the case of semiconductors and insulators so that $\theta(\Delta E-E_{g})=0$ if $\Delta E < E_{g}$ and $\theta(\Delta E-E_{g})=1$ if  $\Delta E > E_{g}$. Although the dependence of $\hslash \omega_{0i\textbf{q}}$ on $\textbf{q}$ is generally unknown, eq.~\eqref{hcortada} is generally accepted using $\zeta$ as an adjustable parameter. The value of $\zeta$ is related to the effective mass of the electrons, so that for free electrons $\zeta=1$ and for insulators with flat bands $\zeta=0$. The calculated fitting parameters of energy loss function for some pure elements and oxides are stored in an open online database \cite{prieto2006electron, sun2017calculations}. The parameters used to model the energy loss function of $h$-BN are listed in Table \ref{tab:template2}. 
\begin{figure}[h] 
    \begin{subfigure}[t]{0.5\textwidth}
	 \centering
 		\includegraphics[width=\columnwidth]{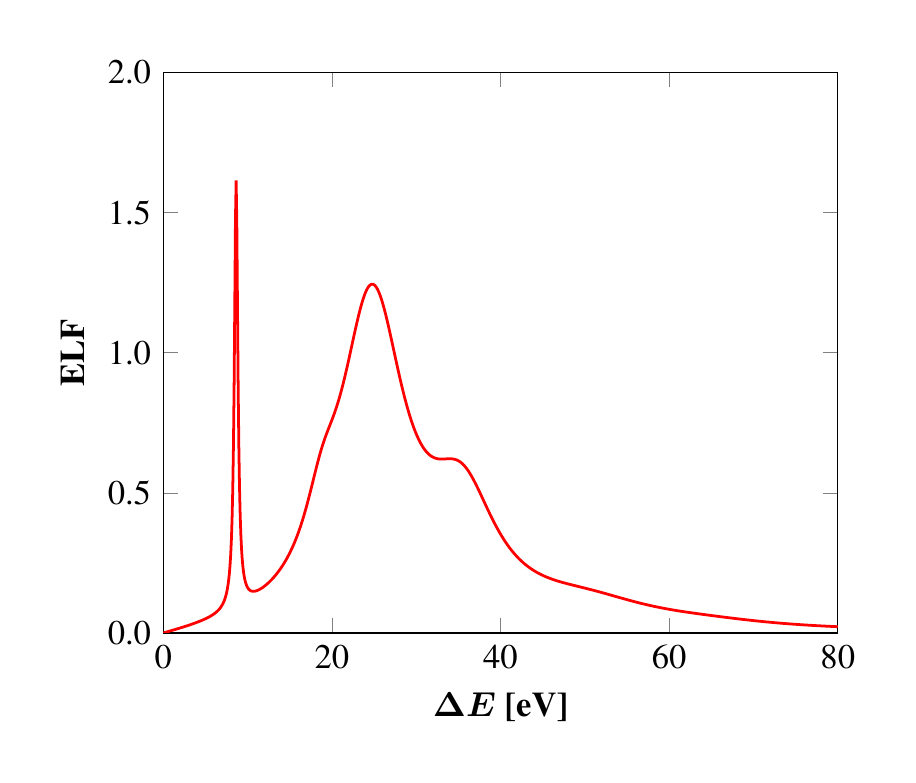}
		\caption{\label{fig:figure2a}}
    \end{subfigure}	
     \begin{subfigure}[t]{0.5\textwidth}
	 \centering
		\includegraphics[width=\columnwidth]{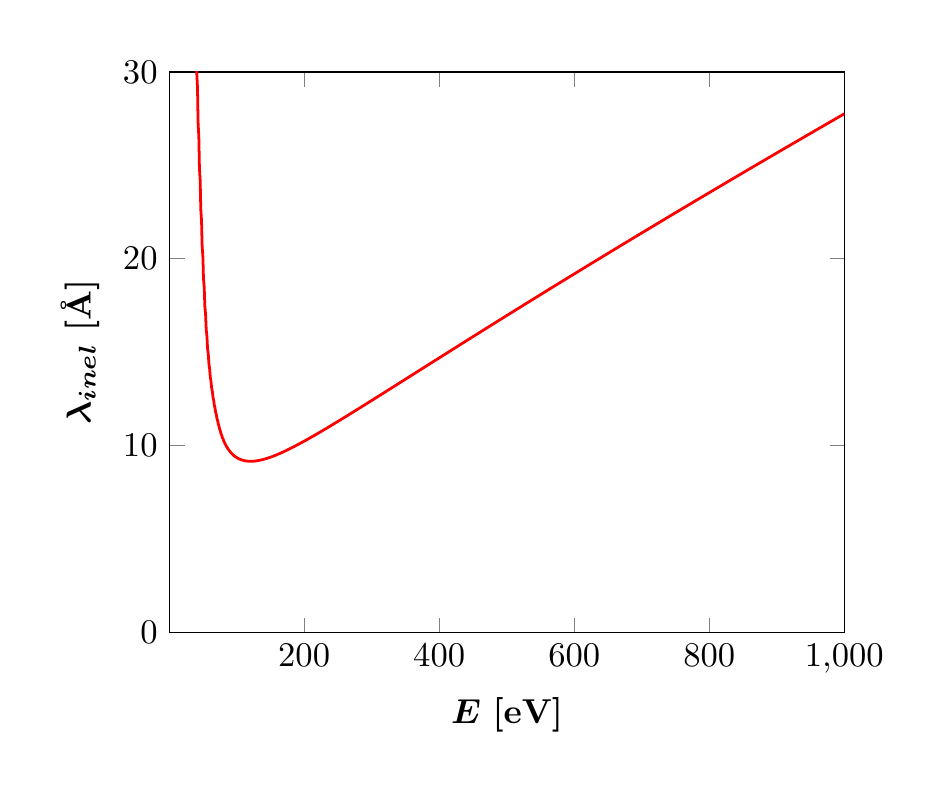}
		\caption{\label{fig:figure2b}}
     \end{subfigure}
\caption{(a) Energy loss function of $h$-BN in the optical limit. (b) Inelastic mean free path of $h$-BN as a function of primary electron energy.}
\label{fig:figure2}	
\end{figure}

To obtain ELF($\textbf{q}$,$\Delta E$) from ELF(0,$\Delta E$) we use Ashley's model \cite{ashley1988interaction} by which the electron differential inelastic scattering cross section can be defined as
\begin{equation}
\frac{d\sigma_{inel}(E,\Delta E)}{d\Delta E} = \frac{me^{2}}{2\pi \hslash^{2}NE} \operatorname{Im}\bigg[-\frac{1}{\varepsilon(0,\Delta E)}\bigg] S\Big(\frac{\Delta E}{E}\Big)
\end{equation}
Here $m$ is the electron mass, $e$ the electron charge, $N$ the number of molecules per unit volume in the target, $E$ the electron energy, and $\Delta E$ the energy transfer. The function $S(x)$ takes the form:
\begin{equation}
S(x) = (1-x)\ln \frac{4}{x}-\frac{7}{4}x+x^{3/2}-\frac{33}{32}x^{2}
\end{equation}
so that the inelastic scattering cross section $\sigma_{inel}(E)$ for the electron-electron interactions can be written as
\begin{equation}
\begin{aligned}
\sigma_{inel}(E) & = \int_{W_{min}}^{W_{max}} \frac{d\sigma_{inel}(E,\Delta E)}{d\Delta E}d\Delta E \\
& = \frac{\lambda_{inel}^{-1}}{N} = \frac{me^{2}}{2\pi \hslash^{2}NE}\int_{W_{min}}^{W_{max}} \operatorname{Im}\bigg[-\frac{1}{\varepsilon(0,\Delta E)}\bigg] S\Big(\frac{\Delta E}{E}\Big) d\Delta E
\end{aligned}
\end{equation}
$W_{min}$ is set to zero for conductors and to the bandgap energy for semiconductors and insulating materials; $W_{max} = E/2$ is the maximum energy transfer.
\begin{equation}
\lambda_{inel}^{-1}(E) = \frac{me^{2}}{2\pi \hslash^{2}E}\int_{W_{min}}^{W_{max}} \operatorname{Im}\bigg[-\frac{1}{\varepsilon(0,\Delta E)}\bigg] S\Big(\frac{\Delta E}{E}\Big) d\Delta E
\end{equation}
Figure \ref{fig:figure2a} is the ELF of $h$-BN in the optical limit; Figure \ref{fig:figure2b} is the inelastic mean free path of $h$-BN as a function of primary electron energy.

For the stopping power $SP = -dE/ds$ , the following expression is used \cite{dapor2014transport}:
\begin{equation}
-\frac{dE}{ds} = \frac{me^{2}}{\pi \hslash^{2}E}\int_{0}^{W_{max}} \operatorname{Im}\bigg[-\frac{1}{\varepsilon(0,\Delta E)}\bigg] G\Big(\frac{\Delta E}{E}\Big) \Delta E d\Delta E
\end{equation}
where
\begin{equation}
G(x) = \ln\frac{1.166}{x}-\frac{3}{4}x-\frac{x}{4}\ln \frac{4}{x}+\frac{1}{2}x^{3/2}-\frac{x^{2}}{16}\ln \frac{4}{x}-\frac{31}{48}x^{2}.
\end{equation}

In order to find the energy loss, $W$, of an inelastic collision of an incident electron with kinetic energy $E$, it is necessary to calculate the function $P_{inel}(W,E)$ providing the fraction of electrons losing energy less than or equal to $W$. 
\begin{equation}
R = P_{inel}(W,E) = \frac{1}{\sigma_{inel}}\int_{0}^{W}\frac{d\sigma_{inel}}{d\Delta E}d\Delta E
\end{equation}
where $R$ is a random number uniformly distributed in the range (0,1].

\begin{table}[htbp]
\centering
\caption{Parameters used to model the energy loss function of $h$-BN.}
\begin{tabular}{l c c c}
\toprule
$h$-BN & $\hslash \omega_{0i}$ $[\rm eV]$ & $A_{0i}$ $[\rm eV^{2}]$ & $\gamma_{0i}$ $[\rm eV]$ \\
($\zeta$ = 0.05) & & & \\
\midrule
1 & 8.65 & 6.6 & 0.5 \\
2 & 19.0 & 18.0 & 5.0 \\
3 & 25.2 & 270.0 & 9.5 \\
4 & 35.8 & 140.0 & 10.0 \\
5 & 51.0 & 80.0 & 20.0 \\
6 & 65.0 & 20.0 & 20.0 \\
\bottomrule
\end{tabular}
\label{tab:template2}
\end{table}

\subsection{Phonon Excitation}
At low energies, when $E$ does not exceed two or three times the value of the bandgap $E_{g}$, an electron has a high likelihood of interacting with the lattice vibrations. The interaction of a quasi-free electron with the longitudinal optical (LO) phonons in a polar medium can be treated by Fr{\"o}hlich's perturbation theory \cite{frohlich1954electrons}. The interaction with the lattice is accompanied by the creation or by the absorption of a phonon. For the optical branch, it is reasonable to ignore the dispersion relation of the longitudinal phonon and to characterize it by the unique frequency $\omega_{LO}$. Then, an electron with  energy $E$ has a probability per unit of path length to create a phonon of frequency $\omega$ (thus losing an  energy $\Delta E = \hslash\omega$) given by
\begin{equation}
\lambda_{ph}^{-1} = \frac{1}{a_{0}}\bigg[\frac{n(T)+1}{2}\bigg]\bigg[\frac{\varepsilon(0)-\varepsilon(\infty)}{\varepsilon(0)\varepsilon(\infty)})\bigg]\frac{\hslash\omega}{E} \ln \bigg\{\frac{[1+\sqrt{1-\hslash\omega/E}]}{[1-\sqrt{1-\hslash\omega/E}]} \bigg\}
\end{equation}
where $a_{0}$ is the Bohr radius, $k_{B}$ is the Boltzmann constant, $\hslash\omega$ is the electron energy loss (on the order of 0.1 eV), $\varepsilon(0)$ is the static dielectric constant, $\varepsilon(\infty)$ is the high frequency dielectric constant and $n(T) = \frac{1}{e^{\hslash\omega/k_{B}T-1}}$
is the occupation number for the phonon level
at temperature $T$, taken here equal to 300 K. For the present calculation, we assume $\varepsilon(\infty) = 4.5$ and $\varepsilon(0) = 7.1$ and only one LO phonon mode has been considered (with energy $\Delta E = \hslash\omega_{LO} = 0.1$ eV). Since the phonon generation probability is higher than the absorption probability by a factor of about 10, the annihilation of the LO phonons along the electron path is neglected.

The polar scattering angle is given according to Llacer \textit{et al.} \cite{llacer1969electron} by
\begin{equation}
\begin{aligned}
& \cos\theta = \bigg(\frac{E+E'}{2 \sqrt{E E'}}\bigg)(1-B^{R})+B^{R} \\
& B = \frac{E+E'+2\sqrt{E E'}}{E+E'-2\sqrt{E E'}}
\end{aligned}
\end{equation}
where $E$ and $E'$ are the electron energy before and after electron-phonon scattering, respectively. $R$ is also a random number uniformly distributed in the range [0,1].

\subsection{Polaronic Effects}
A low-energy electron moving in an insulating material induces a polarization field that has a stabilizing effect on the moving electron. This phenomenon can be described as the generation of a quasi-particle called polaron. The polaron has a relevant effective mass and mainly consists of an electron (or a hole created in the valence band) with its polarization cloud around it. The polaronic effect is important in the description of low-energy electron transport in insulators as it allows the description of the electron trapping and de-trapping necessary for the investigation of electric current inside insulators and charging-up phenomena. However, cross sections of these channels of electron interaction with matter are scarce in the literature. Here, a semi-empirical formula proposed by Ganachaud and Mokrani \cite{ganachaud1993study,ganachaud1995theoretical} is adopted. They assume that the inverse inelastic mean free path that rules the phenomenon --and which is proportional to the probability for a low-energy electron to be trapped in the ionic lattice-- is given by
\begin{equation}
\lambda_{pol}^{-1}(E)= Ce^{-\eta E}
\end{equation}
where $C$ and $\eta$ are constants depending on the dielectric material. \cite{ganachaud1993study,ganachaud1995theoretical}


\section{Monte Carlo Calculation}
\begin{figure}[h] 
\begin{center}
\includegraphics{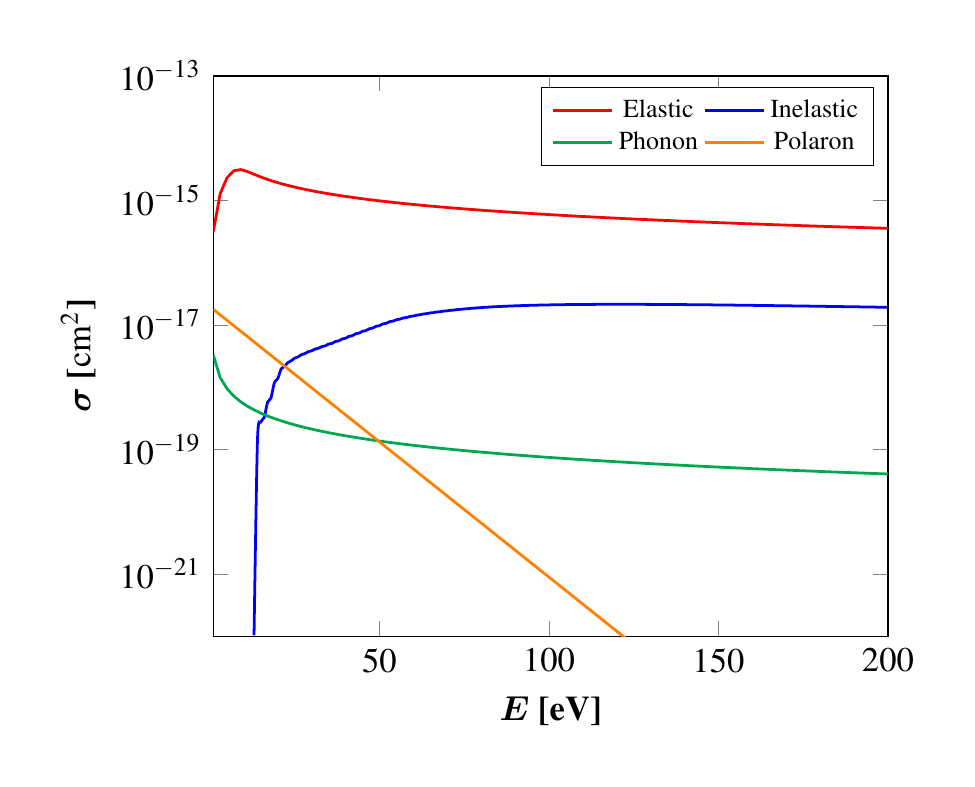}
\caption{Plot of the various scattering cross sections of the processes considered here as a function of primary electron energy.}
\label{fig:figure3}
\end{center}
\end{figure}

The probability for any given scattering to occur is in proportion to its cross section. Thus, specifying the cross section for a given reaction is a proxy for stating the probability that a given scattering process will occur. Figure \ref{fig:figure3} shows various scattering cross sections as a function of primary electron energy.

The stochastic process for multiple scattering is assumed to follow Poisson statistics. If $R$ is a random number uniformly distributed in the interval $(0,1]$, the step length $\Delta s$ is given by
\begin{equation}
\Delta s = -\lambda_{T} \ln R,
\end{equation}
where $\lambda_{T}$ is the electron mean free path, given by
\begin{equation}
\frac{1}{\lambda_{T}} = \frac{1}{\lambda_{el}}+\frac{1}{\lambda_{inel}}+\frac{1}{\lambda_{ph}}+\frac{1}{\lambda_{pol}}.
\end{equation}
The procedure to sample different scattering events follows the sequence:
\begin{equation}
\begin{aligned}
&0 < R \leq \frac{1/\lambda_{el}}{1/\lambda_{T}} \Longrightarrow \text{elastic scattering} \\
&\frac{1/\lambda_{el}}{1/\lambda_{T}} < R \leq \frac{1/\lambda_{el}+1/\lambda_{inel}}{1/\lambda_{T}} \Longrightarrow \text{inelastic scattering} \\
&\frac{1/\lambda_{el}+1/\lambda_{inel}}{1/\lambda_{T}} < R \leq \frac{1/\lambda_{el}+1/\lambda_{inel}+1/\lambda_{ph}}{1/\lambda_{T}} \Longrightarrow \text{phonon excitation} \\
&\frac{1/\lambda_{el}+1/\lambda_{inel}+1/\lambda_{ph}}{1/\lambda_{T}} < R \leq 1 \Big( \equiv \frac{1/\lambda_{el}+1/\lambda_{inel}+1/\lambda_{ph}+1/\lambda_{pol}}{1/\lambda_{T}}\Big) \\
& \Longrightarrow \text{polaron generation}
\end{aligned}
\end{equation}

In this fashion, events are sequentially sampled and executed, conforming effective trajectories that are tracked until an electron reaches the surface with an energy larger than the workfunction or until the electron is thermalized inside the material. Electrons that escape the surface are tallied and the net yield s computed as the ration of the number of escaped electrons relative to the total number of primary trajectories generated.

Our model accounts for the main physical processes on which the secondary electron emission of metal oxides depends. However, several parameters appear in the empirical laws we have proposed. The physical meaning of these parameters is quite clear but their values can be found only by comparing the simulation results to the experimental measurements. 
Preliminary calculations have allowed us to estimate what can be considered as a set of reference parameters. These values will be further varied to check the influence of their choice, particularly on the secondary electron emission yields. The parameters for $h$-BN in present calculation can be found in Table \ref{tab:template1}.

\begin{table}[h]
	\centering
	\caption{Parameters for $h$-BN in present calculation. Among these, $\alpha_{c}$, $W_{ph}$, $C$ and $\eta$ are free parameters. }
	\begin{tabular}{l c c c c c c c c}
		\toprule
		& \multicolumn{8}{c}{Parameters} \\
		\cmidrule(l){2-9}
		$h$-BN & $\alpha_{c}$ & $E_{g}$ [eV] & $\varepsilon(\infty)$ & $\varepsilon(0)$ & $W_{ph}$ [eV] & $C~[\rm nm^{-1}]$ & $\eta~[\rm eV^{-1}]$ & $\chi$ [eV] \\ 
		\midrule
		& 0.5 & 5.2 & 4.10 & 5.09 & 0.1 & 1.0 & 0.1 & 4.5 \\
		\bottomrule
	\end{tabular}
		\label{tab:template1}
\end{table}

\section{Results}
\subsection{Flat Surfaces}
The total secondary electron yield for ideally-flat $h$-BN surfaces is calculated for incident angles of 0$^{\circ}$, 15$^{\circ}$, 30$^{\circ}$, 45$^{\circ}$, 60$^{\circ}$, 75$^{\circ}$ and 89$^{\circ}$ measured off the surface normal, and incident energies in the range 50-1000 eV. In this work, the typical number of primary particles simulated ranges between $10^{4}$ and  $10^{5}$, which generally results in statistical errors around 3$\%$.

Experimental data on $h$-BN secondary electron emission yield is scarce and subjected to high uncertainty, so first we test our models on a material system such as SiO$_{2}$ for which such data exist. This is a first step aimed at validating our codes before the study of BN. The results for SiO$_{2}$ are given in the Appendix, which convincingly demonstrates the validity of our models.
On this basis, the simulations of 50 eV-1 keV electron irradiation on flat $h$-BN surfaces are performed. 

\begin{figure}[h]
    \centering
    \begin{subfigure}[t]{0.5\textwidth}
        \centering
        \includegraphics[width=\columnwidth]{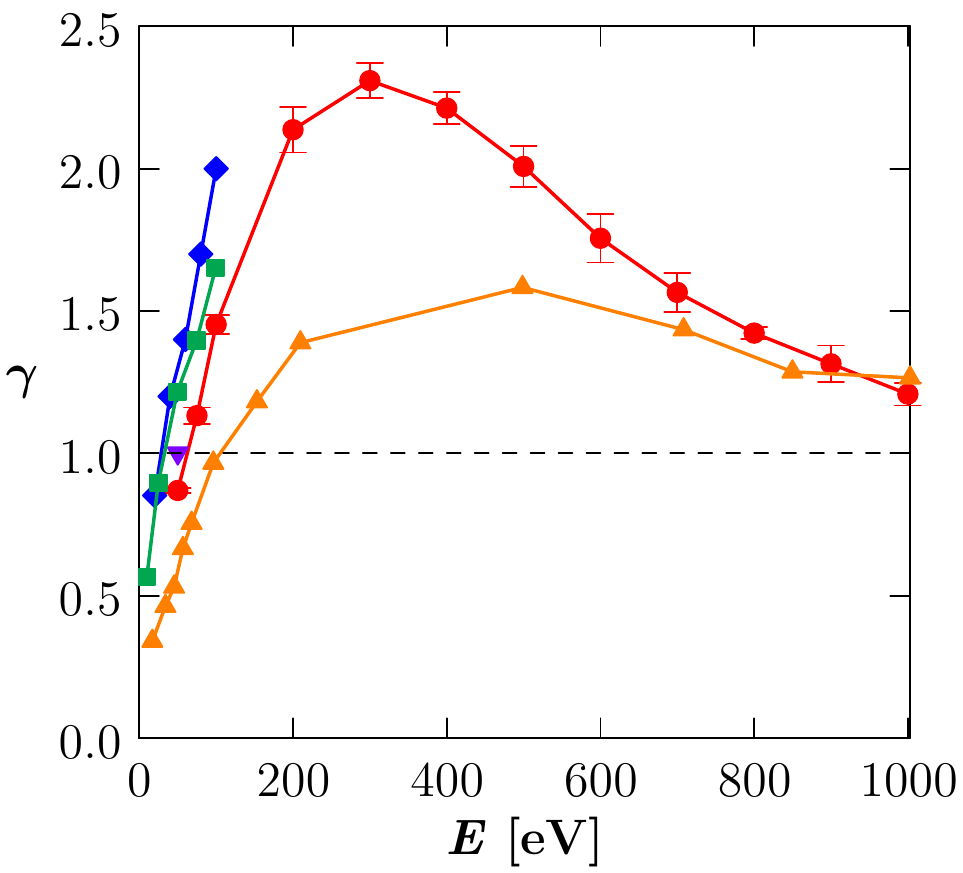}
        \caption{\label{fig:figure4a}}
    \end{subfigure}%
    \begin{subfigure}[t]{0.5\textwidth}
        \centering
        \includegraphics[width=\columnwidth]{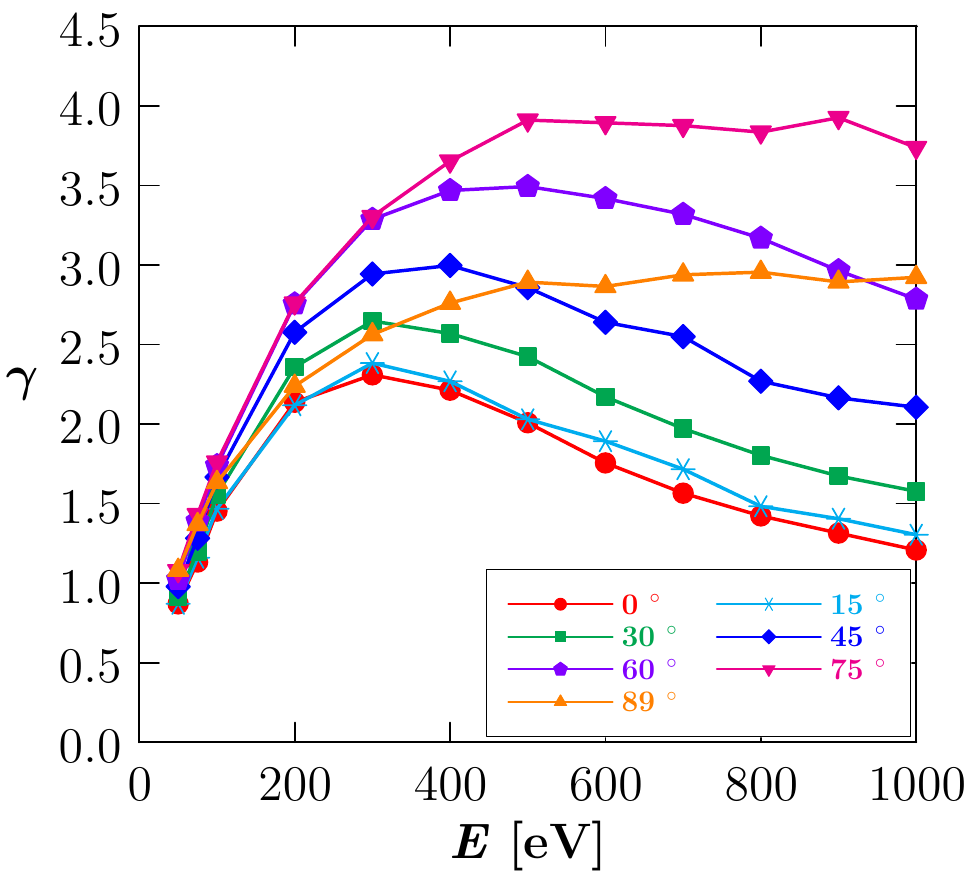}
        \caption{\label{fig:figure4b}}
    \end{subfigure}
    \caption{(a) Total SEE yield from smooth  $h$BN surface as a function of primary electron energy for electrons incident at 0$^{\circ}$. $\bullet$ = this work; $\blacktriangledown$ = Dawson (1966) \cite{dawson1966secondary}; $\blacklozenge$ = ONERA (1995) \cite{bugeat1995development}; $\blacksquare$ = PPPL (2002) \cite{dunaevsky2003secondary}; $\blacktriangle$ = Christensen (2016) \cite{christensen2016instrumentation}.  (b) Total SEE yield from an ideally-flat $h$BN as a function of primary electron energy, for electrons incident at 0$^{\circ}$, 15$^{\circ}$, 30$^{\circ}$, 45$^{\circ}$, 60$^{\circ}$, 75$^{\circ}$ and 89$^{\circ}$.}
    \label{fig:figure4}
\end{figure}

At low energy regime, the simulation results are found to agree reasonably well with experimental data, which corresponds exactly to normal operating conditions ($<$100eV) of Hall thrusters. The agreement is slightly worse in the intermediate energy regime, which we rationalize in terms of the charging effect of insulators under electron irradiation. Again, at high energy regime, the simulation results converge to the experimental data. Note that under steady state, the total SEE yield at high temperatures should tend to unity. This can be attributed to the onset of a charge gradient due to the existence of holes created by the departure of secondary electrons from the lattice. This charge gradient creates a `shielding' electric field that captures further SEE until charge neutrality is achieved again. Once the material is neutral, the process starts again, leading to an oscillatory steady state that keeps SEE balanced \cite{thomson2003electron}.
This picture can be altered by factors such as radiation-induced conductivity changes, sheath potential modifications, or slowly evolving internal charge distributions. In addition, the roughness of 'real' experimental surfaces compared to the ideally-smooth ones in the model surely plays a role in the comparison. SEE yields as a function of $E$ for all angles of incidence considered here are given in Figure \ref{fig:figure4b}.

\begin{figure}[h]
    \centering
    \begin{subfigure}[t]{0.5\textwidth}
        \centering
        \includegraphics[width=\columnwidth]{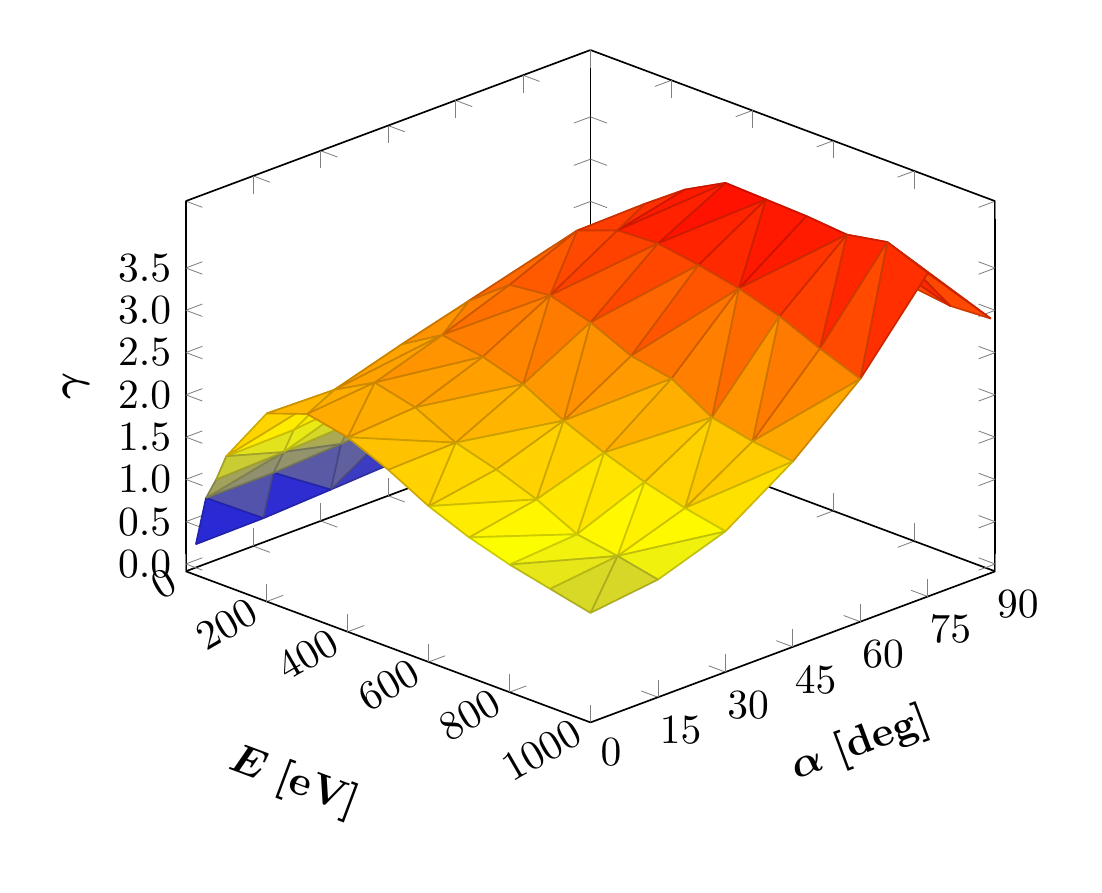}
        \caption{\label{fig:figure5a}}
    \end{subfigure}%
    \begin{subfigure}[t]{0.5\textwidth}
        \centering
        \includegraphics[width=\columnwidth]{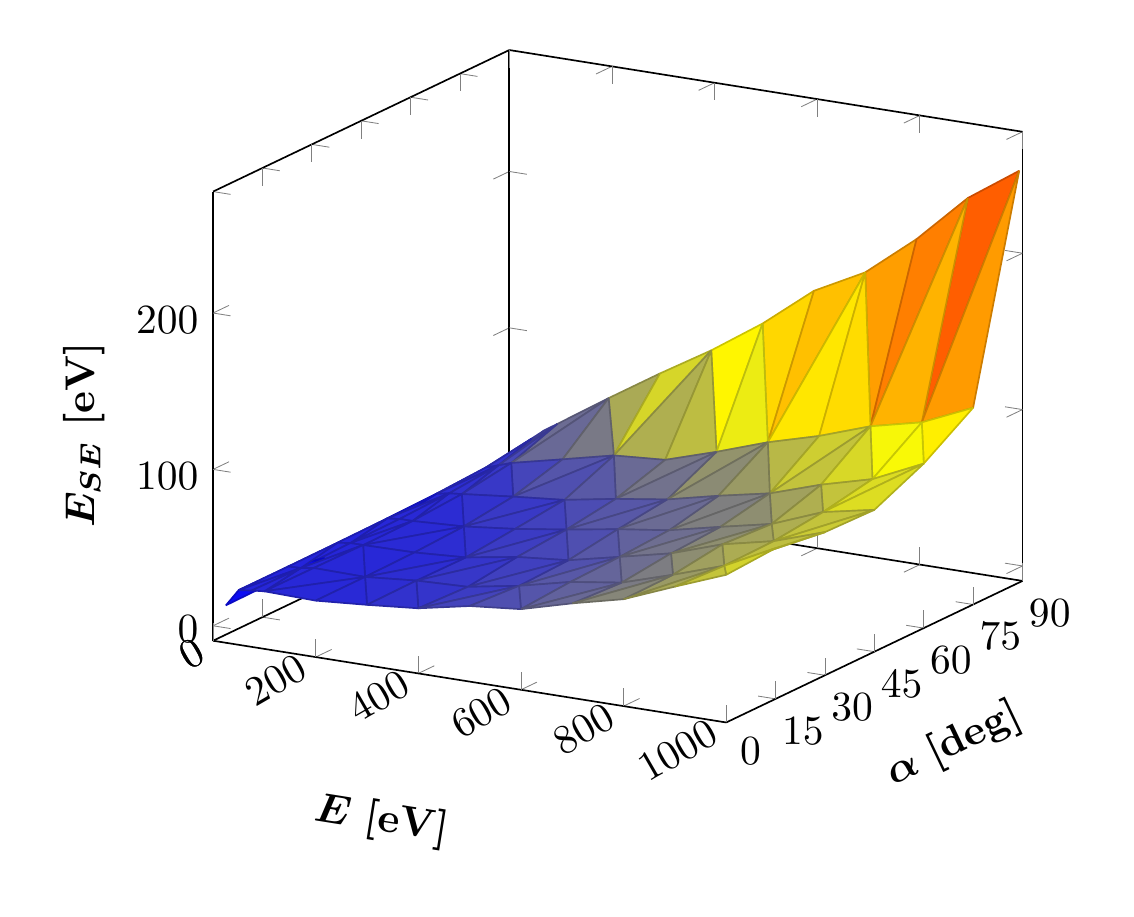}
        \caption{\label{fig:figure5b}}
    \end{subfigure}
    \caption{(a) Surface plot of the total SEE yield from an ideally-flat $h$BN as a function of primary electron energy and angle of incidence. (b) Surface plot of the SEE energy distributions from an ideally-flat $h$BN as a function of primary electron energy and angle of incidence.}
    \label{fig:figure5}
\end{figure}

Surface plots of both the SEE energy distributions and the yields are given in Figures \ref{fig:figure5a} and \ref{fig:figure5b}.  As mentioned earlier, these data will be used in ray-tracing Monte Carlo simulations of SEE in arbitrary surface geometries. The details of the function fitting process can be found in our prior publications \cite{chang2018calculation,alvarado2018monte}. 
The final expressions for the total SEE yield and energy distributions are generated using machine learning software \cite{schmidt2009distilling}:
\begin{equation}
\gamma(E,\alpha) = \begin{cases} 
\begin{aligned}
& 0.0185E+1.53\times 10^{-15}E^{4}+1.53\times 10^{-15}E^{3}+1.2\times 10^{-6}E\alpha^{2}+1.53\times10^{-15}E^{4}\alpha^{3} \\
& 3.915\times 10^{-5}E^{2}-1.4\times 10^{-12}E^{2}\alpha^{4} & 0 \leq E \leq 200\mbox{eV}
\end{aligned} \\ \\
\begin{aligned}
& 3.82+1.54\times 10^{-6}E^{2} \cos(2.48 \alpha)-3.68 \times 10^{4} / (2.63\times 10^{-1}\alpha+E^{2}-E \alpha  \cos^{2}(2.48 \alpha)) \\
& -7.88\times 10^{-4} E -3.37\times 10^{-3} E \cos(2.48\alpha) & 200 \leq E \leq 1000\mbox{eV} 
\end{aligned}
\end{cases}.
\label{corr1}
\end{equation}
\begin{equation}
E_{SE}(E,\alpha) = \begin{cases} 
\begin{aligned}
& 0.73E+3.88\times 10^{-6}\alpha E^{2}+6.60\times 10^{-7}E \alpha^{3}+E^{2}\sin(5.31\times 10^{-8}E^{2}) \\
& -5.12\times 10^{-3}E^{2}-4.22\times 10^{-5}E\alpha^{2} & 0 \leq E \leq 200\mbox{eV}
\end{aligned} \\ \\
\begin{aligned}
& 27.81+7.69\times 10^{-5} E^{2}-1.99\times 10^{5}/(E^{3}\cos(E))+(12.16-1.94E)/(\alpha-97.58) \\
& -3.97\times 10^{-2} E-4.08\times10^{-7}\alpha E^{2} & 200 \leq E \leq 1000\mbox{eV} 
\end{aligned}
\end{cases}.
\label{corr2}
\end{equation}

\subsection{Micro-Architectured Foam Structures}
Next we calculate SEE from microfoam structures with various porosities. The details of these structures are given in our past studies \cite{chang2018calculation,alvarado2018monte}. A finite element reconstruction of the material is used to extract surface elements that may be intersected by electron trajectories. Special algorithms are then used to identify intersections between primary and secondary rays. Daughter rays are generated from parent rates using correlations \eqref{corr1} and \eqref{corr2}.

\begin{figure}[h]
    \centering
    \begin{subfigure}[t]{0.55\textwidth}
        \centering
        \includegraphics[width=\columnwidth]{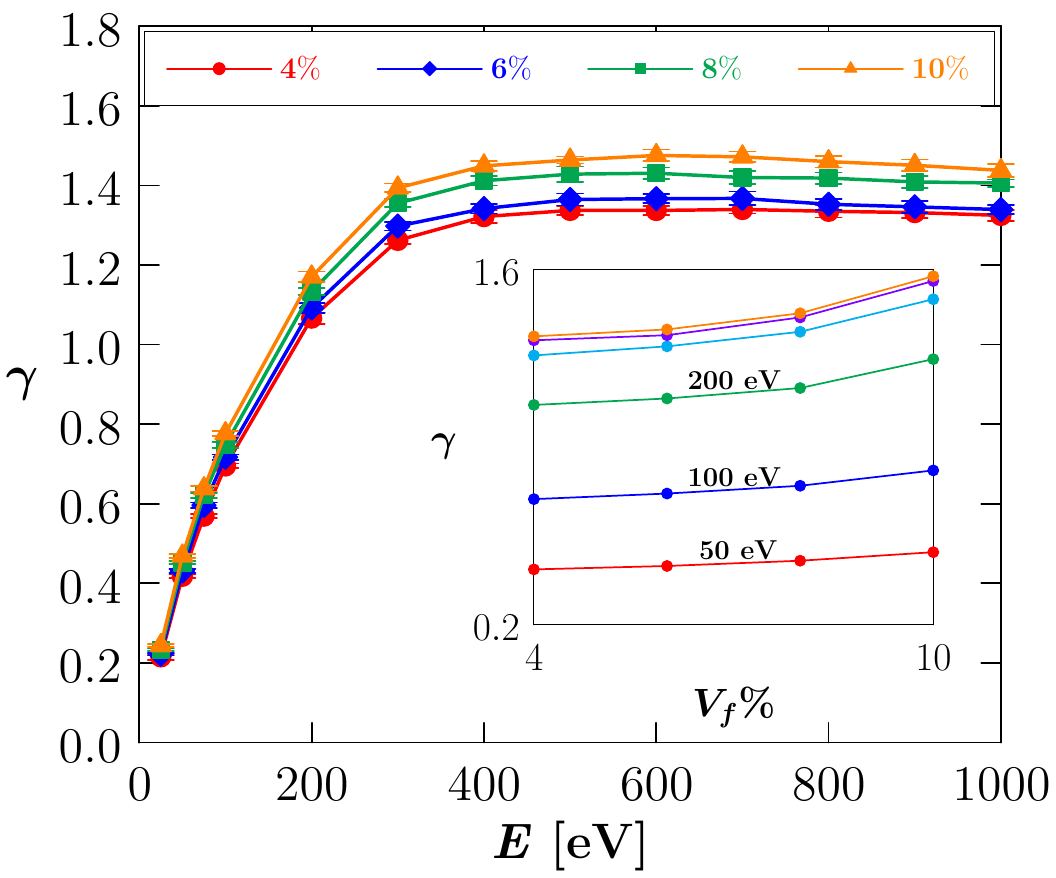}
        \caption{\label{fig:figure6a}}
    \end{subfigure}%
    \begin{subfigure}[t]{0.5\textwidth}
        \centering
        \includegraphics[width=\columnwidth]{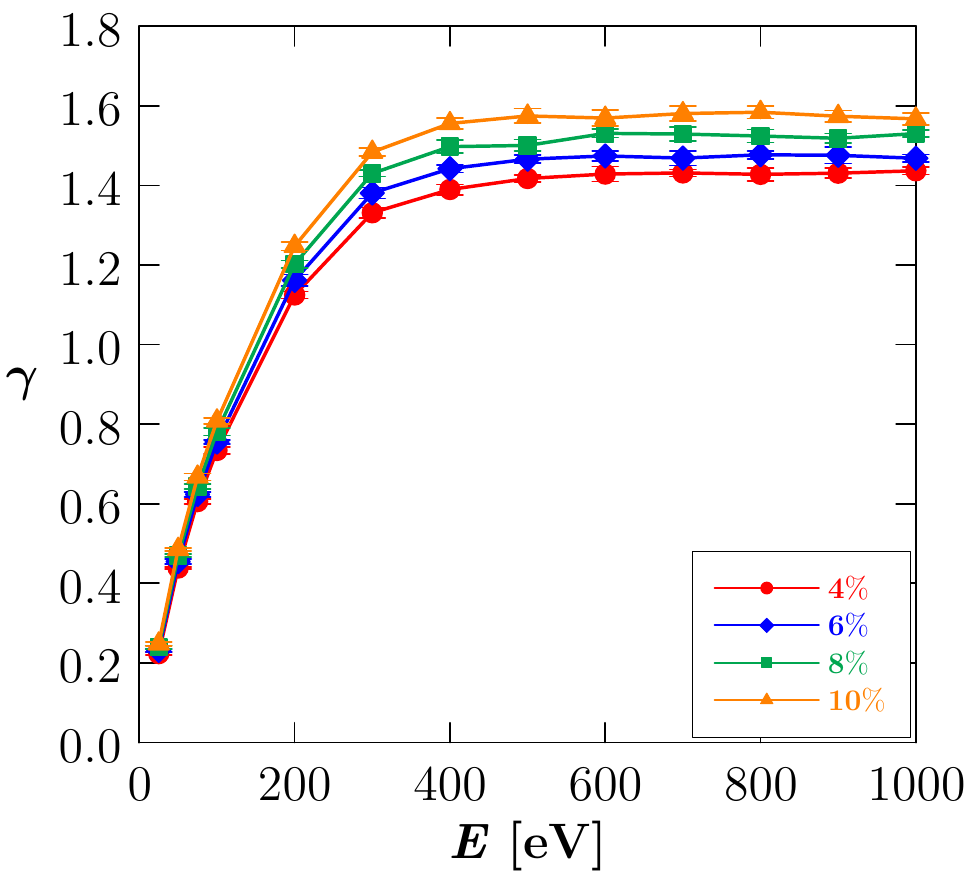}
        \caption{\label{fig:figure6b}}
    \end{subfigure}
    \caption{(a) SEE yield versus electron beam energy for normal incidence to the foam at varying volume-fraction percentages. The inset shows (in increasing order) the dependence of the yield with volume fraction for primary energies equal to 50, 100, 200, 300, 400, 500, and 600 eV. (b) SEE yield versus electron beam energy for random incidence to the foam at varying volume-fraction percentages.}
    \label{fig:figure6}
\end{figure}

Figures \ref{fig:figure6a} and \ref{fig:figure6b} show the secondary electron emission yield for solid volume fractions, $V_f$, of 4, 6, 8, and 10\% in the 50-to-1000-eV energy range for normal and random incidence. The inset to figure \ref{fig:figure6a} shows the dependence of the yield with the material volume fraction at energies of 50, 100, 200, 300, 400, 500, and 600 eV for normal incidence. In the high porosity range explored here the dependence of the SEE yield on $V_{f}$ is clearly linear in the high porosity range explored here. For the sake of comparison, the maximum SEE yield for $V_{f}$ = 4\% (which occurs for $E$ = 700 eV) is approximately 1.3, compared with a value of 2.3 for the flat surface (from Fig.\ \ref{fig:figure4b}). This decrease in SEE yield by about a factor of two is indicative of the potential performance gains that micro-architected surfaces might offer relative to fully dense surfaces.

\section{Discussion and conclusions}

The main objective of our work is to develop reliable physical models of secondary electron emission (SEE), to be applied to the calculation of effective SEE yields in micro-architected surfaces for space propulsion thrusters. The methodology relies on two distinct but complementary elements. The first is an experimentally validated theoretical model of electron scattering in solids. This model is built as a transport Monte Carlo simulator of individual electron trajectories in a solid, capturing the pertinent scattering mechanisms in terms of an interaction differential cross section that is integrated across the relevant energy and angular ranges. These cross sections reflect different elastic and inelastic (e.g. core electrons, valence electrons, polarons, plasmons, etc) scattering mechanisms in each material, which are formulated according to the best available physics. The second element of the methodology is a discretization procedure to represent arbitrary surface geometries in terms of discrete boundary elements. Both modules (physics and geometry) are coupled by way of a raytracing algorithm that captures the intersection of primary electron rays with different boundary elements. The material considered here is BN (used in current thrusters thanks to its low mass density, low thermal expansion, and highly dielectric properties).

Our main findings are that, in the primary electron energy range of interest ($<$100 eV) micro-foam architected surfaces are seen to decrease the SEE yield by about a factor of two. This is already a significant advantage over flat surfaces (or with as-fabricated surface roughness). As well, these micro-foams are seen to suppress the SEE yield peak typically observed at 400 eV of primary energy. Our results have been validated at two different levels.  As mentioned above, the electron scattering model has been compared to experiments in smooth surfaces for both SiO$_2$ and BN. As this is the `physics' building block of the methodology, great care has been placed on ensuring that the parameters of the model are consistent with available measurements in each case. Second, we have compared results for the W micro-foam to measurements in `fuzz' W surfaces (smooth W surfaces pre-exposed to a He plasma), which have a similar degree of porosity and morphology as foams. 
 
Our approach has several advantages. First, it allows us to study any arbitrary surface morphology thanks to the finite element discretization scheme and the raytracing Monte Carlo method to generate electron trajectories. This endows the methodology with an extraordinary versatility, as complex surface geometries such as foams, pillars, fuzz, cells, etc., of any size can be treated in a straightforward manner. As well, our physical scattering model being capable of describing both metals and ceramics, we can now study materials of high relevance for electric propulsion. We believe that this approach will enable a rapid parsing of thruster lining material surface concepts prior to costly development and characterization, to narrow down the parametric space and accelerate materials development.

To summarize, our main findings are:
\begin{enumerate}
\item In the primary electron energy range of interest ($<$100 eV) BN microfoam architected surfaces are seen to decrease the SEE yield by about a factor compared to ideally flat surfaces. 
\item These micro-foams are seen to suppress the SEE yield peak typically observed at 400 eV of primary energy.
\item Our results have been validated by comparison to experiments in smooth surfaces for both SiO$_2$ and BN. 
\end{enumerate}
 
\section*{Acknowledgement}
The authors acknowledge support from the Air Force Office of Scientific Research (AFOSR), through award number FA9550-11-1-0282 with UCLA.	

\bibliography{references}

\appendix
\renewcommand\thefigure{\arabic{figure}}   
 
\label{key}
\section{List of Symbols}
\input{symbols.tex}
\input{SiO2.tex}

\end{document}

%% file: symbols.tex
\begin{tabular}{ l  l }
  $E$ & kinetic energy of the primary electron [eV] \\
  $E_{g}$ & bandgap [eV] \\
  $A$ & atomic weight [g/mol] \\
  $\rho$ & density of the target [g/cm$^{3}$] \\
  $N_{a}$ & Avogadro's number = 6.02 $\times 10^{23}$ \\
  $N=\rho N_{a}/A$ & atomic number density \\
  $Z$ & atomic number \\
  $a_{0}$ & Bohr radius = 0.529 [$\AA$] \\
  $k_{B}$ & Boltzmann constant \\
  $\Delta E$ & energy loss of primary electron [eV] \\
  $q$ & momentum transfer \\
  $\hslash$ & reduced Planck constant \\
  $\omega$ & phonon frequency \\
  $\chi$ & electron affinity [eV] \\
  $\varepsilon(0)$ & static dielectric constant \\
  $\varepsilon(\infty)$ & high frequency dielectric constant \\
  $n$ & refractive index \\
  $k$ & extinction coefficient \\
  $\alpha$ & incident angle of primary electron [deg] \\
  $\gamma$ & secondary electron emission yield \\
  $R_{c}$ & cut-off function \\
  $\sigma_{el}$ & elastic scattering cross section [$\AA^{2}$] \\
  $\sigma_{inel}$ & inelastic scattering cross section [$\rm \AA^{2}$] \\
  $\sigma_{ph}$ & phonon excitation cross section [$\rm \AA^{2}$] \\
  $\sigma_{pol}$ & polaron cross section [$\rm \AA^{2}$] \\
  d$\sigma$/d$\Omega$ & differential scattering cross section with respect to direction \\
  d$\sigma$/dE & differential scattering cross section with respect to energy \\
  $\lambda_{el}$ & elastic mean free path [$\rm \AA$] \\
  $\lambda_{inel}$ & inelastic mean free path [$\rm \AA$] \\
  $\lambda_{ph}$ & phonon excitation mean free path [$\rm \AA$] \\
  $\lambda_{pol}$ & polaron excitation mean free path [$\rm \AA$] \\
  $\lambda_{T}$ & total mean free path [$\rm \AA$] \\
  $\theta$ & polar scattering angle of the primary electron [deg] \\
  $\vartheta$ & polar scattering angle of the secondary electron [deg] \\
  $\phi$ & azimuthal scattering angle of the primary electron [deg] \\
  $\varphi$ & azimuthal scattering angle of the secondary electron [deg] \\
  \end{tabular}

%% file: SiO2.tex
\begin{figure}[h] 
    \begin{subfigure}[t]{0.5\textwidth}
	 \centering
 		\includegraphics[width=\columnwidth]{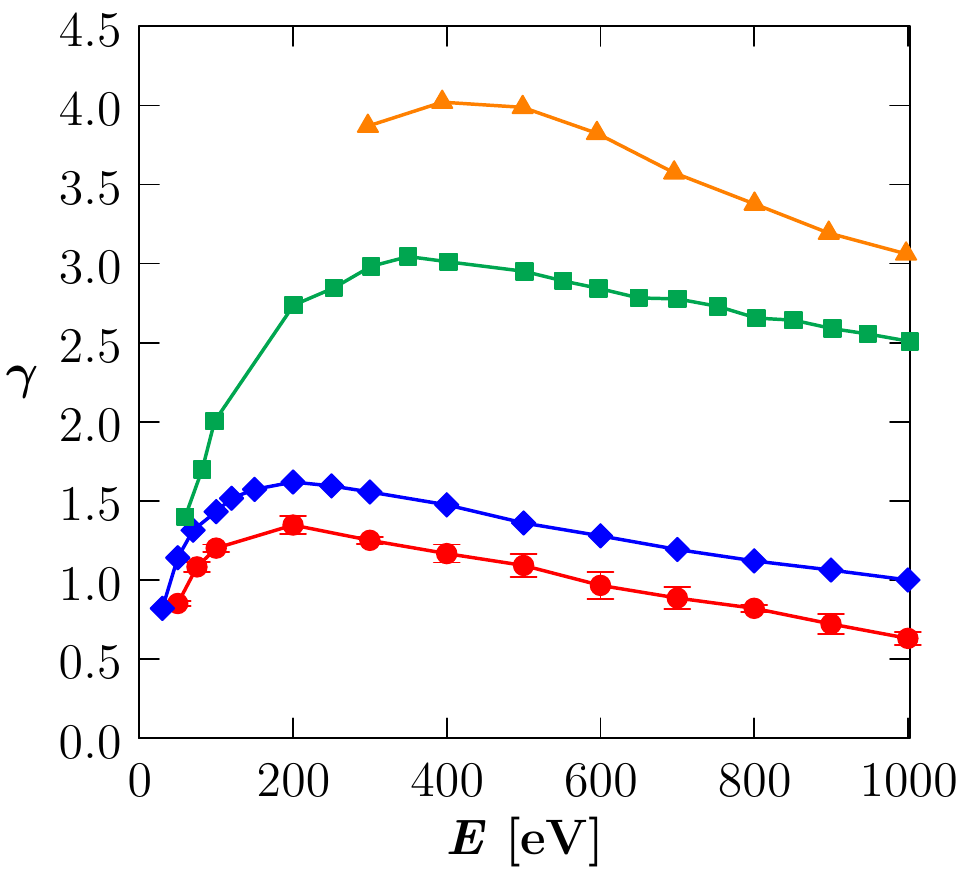}
		\caption{\label{fig:figure5a}}
    \end{subfigure}	
     \begin{subfigure}[t]{0.5\textwidth}
	 \centering
		\includegraphics[width=\columnwidth]{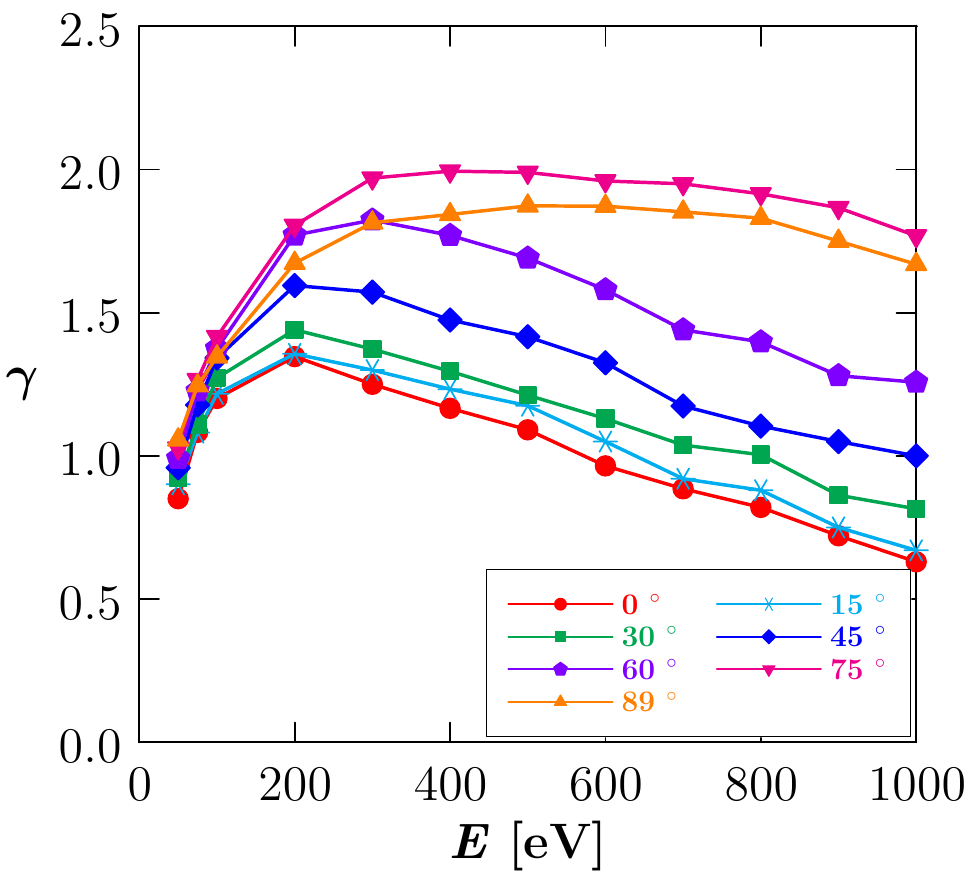}
		\caption{\label{fig:figure5b}}
     \end{subfigure}
\caption{(a) Total SEE yield from smooth SiO$_{2}$ surface as a function of primary electron energy for electrons incident at 0$^{\circ}$. $\bullet$ = this work; $\blacklozenge$ = Dionne (1975) \cite{dionne1975origin}; $\blacksquare$ = Barnard (1977) \cite{barnard1997measurements}; $\blacktriangle$ = Yong (1998) \cite{yong1998determination}.  (b) Total SEE yield from an ideally-flat SiO$_{2}$ as a function of primary electron energy, for electrons incident at 0$^{\circ}$, 15$^{\circ}$, 30$^{\circ}$, 45$^{\circ}$, 60$^{\circ}$, 75$^{\circ}$ and 89$^{\circ}$.}
\label{fig:figure7}	
\end{figure}